\documentstyle[11pt]{article}
\def\fnote#1#2{\begingroup\def\thefootnote{#1}\footnote{#2}\addtocounter
{footnote}{-1}\endgroup}
\def\BM#1{\mbox{\boldmath{$#1$}}}

\begin{document}

\hfill{UTTG-21-13 }

\vspace{36pt}

\begin{center}
{\large {\bf {Quantum Mechanics Without State Vectors}}}

\vspace{36pt}
Steven Weinberg\fnote{*}{Electronic address:
weinberg@physics.utexas.edu}\\
{\em Theory Group, Department of Physics, University of
Texas\\
Austin, TX, 78712}

\vspace{30pt}

\noindent
{\bf Abstract}
\end{center}

It is proposed to give up the description of physical states in terms of ensembles of state vectors with various probabilities, relying instead solely on the density matrix as the description of reality.  With this definition of a physical state,  even in entangled states nothing that is done in one isolated system can instantaneously effect the physical state of a distant isolated system.  This change in the description of physical states opens up a large variety of new ways that the density matrix may transform under various symmetries, different from the unitary transformations of ordinary quantum mechanics.  Such new transformation properties have been explored before, but so far only for the symmetry of time translations into the future, treated as a semi-group.  Here new transformation properties are  studied for general symmetry transformations forming groups, rather than semi-groups.  Arguments are given that such symmetries should act on the density matrix as in ordinary quantum mechanics, but loopholes are found for all of these arguments.
\noindent

\vfill

\pagebreak

\begin{center}
{\bf I. A MODEST PROPOSAL}
\end{center}

Two unsatisfactory features of quantum mechanics  have bothered physicists for decades.  The first is the difficulty of dealing with measurement.   The unitary deterministic evolution of the state vector in quantum mechanics cannot convert a definite initial state vector to an ensemble of eigenvectors of the measured quantity with various probabilities.  Here we seem to be faced with nothing but bad choices.  The Copenhagen interpretation[1] assumes  a mysterious division between the microscopic world governed by quantum mechanics and a macroscopic world of apparatus and observers that obeys classical physics.  If instead we take the wave function or state vector seriously as a description of reality, and suppose that it evolves unitarily according to the deterministic time-dependent Schr\"{o}dinger equation, we are inevitably led to a many-worlds interpretation[2], in which all possible results of any measurement are realized.     To avoid both the absurd dualism of the Copenhagen interpretation and   the endless creation of inconceivably many branches of history of the many-worlds approach, some physicists adopt an instrumentalist position, giving  up on any realistic interpretation of the wave function, and regarding it as only a source of predictions of probabilities, as in the decoherent histories approach[3].

The other problem with quantum mechanics arises from entanglement[4].  In an entangled state in ordinary quantum mechanics  an intervention in the state vector affecting one part of a system can instantaneously affect the state vector  describing a distant isolated part of the system.  It is true that in ordinary quantum mechanics no measurement in one subsystem can reveal what measurement was done in a different isolated subsystem, but  the susceptibility of the state vector  to instantaneous change from a distance  casts doubts on its physical significance.  

Entanglement is much more of a problem in some modifications of quantum mechanics that are intended to resolve the problem of measurement, such as  the general nonlinear stochastic evolution studied in [5].  It is difficult in these theories even to formulate what we mean by isolated subsystems, much less to prevent instantaneous communication between them[6,7].  Polchinski[7] has shown that unless nonlinearities are constrained to depend only on the density matrix,   such modified versions of quantum mechanics even allow communication between the different worlds of the many-worlds description of quantum mechanics.

The problem of instantaneous communication between distant isolated systems has been nicely summarized in a theorem of Gisin[6].  It states that in a system consisting of two  isolated subsystems {\em I} and {\em II}, with a prescribed density matrix $\rho^I$ for subsystem ${\em I}$, it is always possible in a suitable entangled state of the two subsystems to make measurements on subsystem $II$ that put subsystem ${\em I}$ in {\em any} set of states $\Psi^I_r$ (not  necessarily orthogonal) with probabilities $P_r$, provided only that $\sum_r P_r
\Lambda^I_r=\rho^I$, where $\Lambda^I_r$ is the projection operator on the state  $\Psi^I_r$.    

Since any statement that a system is in an ensemble of states with definite probabilities can thus be changed instantaneously by a measurement at an arbitrary distance, keeping only the density matrix fixed, it seems reasonable to infer that such statements are meaningless, and that only the density matrix has meaning.  That is, it  seems  worth considering yet another interpretation of quantum mechanics:  The density matrix rather than the state vector or wave function  is to be taken as a description of reality.  

Taking the density matrix as the description of reality is very different from giving the same status to an ensemble of  state vectors with various probabilities, because the density matrix contains much less information.    If we know that a system is in any one of a number of  states $\Psi_r$, with  probabilities $P_r$, then we know that the density matrix is $\rho=\sum_r P_r\Lambda_r$, where $\Lambda_r$ is the projection operator on state  
$\Psi_r$, but this does not work in reverse.  As is well known, for a given density matrix $\rho$ there are any number of ensembles of not necessarily orthogonal or even independent state vectors and their probabilities that give the same density matrix.  (An exception is discussed in Section II.)      The density matrix is of course a Hermitian operator on  Hilbert space, a  vector space.  In speaking of  ``quantum mechanics without state vectors''  I mean only that a statement that a system is in any one of various state vectors with various probabilities is to be regarded as having no meaning, except for what it tells us about the density matrix.  

For example, suppose the density matrix of a spin $1/2$ particle, in a basis provided by states with the north-component of spin equal to $+1/2$ or $-1/2$, takes the form
$$ \rho=\left(\begin{array}{cc} 0.69 & 0.17 \\ 0.17 & 0.31\end{array}\right)\;.$$
By diagonalizing this matrix, we might conclude that 
 this particle has a 75\% probability of being in a pure state with spin pointing northeast and a 25\% probability of being in an orthogonal pure state with spin pointing southwest.  But we get the same density matrix if the particle has a 50\% probability of being in a pure state with spin pointing north, a 15\% probability of being in a pure state with spin pointing south, and a 35\% probability of being in a pure state with spin pointing east.    These ensembles sound different, but in fact they are indistinguishable.  Indeed, they had better be indistinguishable, because otherwise we could communicate instantaneously at an arbitrary distance by acting on a distant isolated system with which this particle's state vector is entangled so as to change the  spin states from the first to the second ensemble.  It is better just to specify the density matrix,
and give up its description in terms of an ensemble of state vectors with various probabilities.

If the density matrix is not to be defined in terms of ensembles of state vectors, then what is it?  We may define it by  postulating a physical interpretation:   The average value $\overline{A}$ of any physical quantity represented by an Hermitian operator $A$ is ${\rm Tr}(A\rho)$, which since it applies also to powers of $A$ allows us to find from the density matrix the probability distribution  for values of $A$.  (These may be regarded as objective probabilities, independent of whether or not anything is actually being measured.)  This postulate leads to all the  properties of the density matrix that are usually derived from its interpretation in terms of an ensemble of states with various probabilities.  The density matrix must be Hermitian in order that ${\rm Tr}(A\rho)$ should be real for an arbitrary Hermitian operator $A$.  The density matrix must have unit trace in order that ${\rm Tr}(\alpha\rho)=\alpha$ for any c-number $\alpha$.  The density matrix must be positive in order that ${\rm Tr}(A\rho)$ should be positive for any positive Hermitian operator $A$.  Also, a physical quantity represented by a Hermitian operator $A$ will have a definite value $\alpha$ (in the sense that the mean value of $A^n$ is $\alpha^n$ for all integers $n$) if and only if $A\rho=\alpha\rho$.

It may seem like a mere matter of language to say that it is the  density matrix rather than  an ensemble of state vectors with various probabilities that should be taken as the description of a physical system.  Already many studies of the interpretation of quantum mechanics and of quantum information theory are  based on the density matrix rather than the state vector, without needing a new interpretation of quantum mechanics.  What difference does it make?  

There is one big difference, that is our chief concern in this paper. Giving up the definition of the density matrix in terms of state vectors opens up a much larger variety of ways that the density matrix might respond to various symmetry transformations.  In ordinary quantum mechanics, a symmetry transformation takes a density matrix $\rho$ into $U\rho U^\dagger$, where $U$ is a unitary (or, for time-reversal, antiunitary) operator belonging to one of the representations of the symmetry group.  This is certainly not the only way that an Hermitian matrix could transform.    For instance, we may consider a system with an $SU(3)$ symmetry and  a Hilbert space of three-dimensions, in which the density matrix transforms under $SU(3)$ as the reducible representation ${\BM 3}+\overline{\BM 3}+{\BM 1}+{\BM 1}+{\BM 1}$.  In a suitable basis, we would have
$$\rho=\left(\begin{array}{ccc}a_1 & b_3 & b_2^* \\ b_3^* & a_2 & b_1 \\ b_2 & b_1^* & a_3 \end{array}\right)\;,$$
where  under $SU(3)$ the real diagonal elements $a_n$ transform as singlets, with $a_1+a_2+a_3=1$, and the triplet $(b_1,b_2,b_3)$ transforms as the representation ${\BM 3}$.
This $SU(3)$ transformation of $\rho$  cannot be put in the form $\rho\mapsto U\rho U^\dagger$ required in ordinary quantum mechanics, because if the $3\times 3$ matrix $U$  belonged to the representations ${\BM 3}$ or $\overline{\BM 3}$ then $\rho$ would transform as  ${\BM 3}\times\overline{\BM 3}= {\BM 1}+{\BM 8}$, not ${\BM 3}+\overline{\BM 3}+{\BM 1}+{\BM 1}+{\BM 1}$.  (The other possibility is that $U$  belongs to the representation ${\BM 1}+{\BM 1}+{\BM 1}$, in which case $\rho$ would transform as a sum of singlets, again not including ${\BM 3}+\overline{\BM 3}$.)  The question of the positivity of a density matrix transforming in this way is discussed in Section VII.

The possibility of an unusual transformation of the density matrix has been widely considered, but up to now I believe only for the symmetry of time-translation.  In this case it is known that the evolution of the density matrix with time is in general governed by a first-order linear differential equation, such as the Lindblad equation[8] (given here in Section VIII), different from what is found in ordinary quantum mechanics.  The Lindblad equation is commonly used to study open systems in ordinary quantum mechanics, with the effects of the environment integrated out, but it has also been  used to deal with the problem of measurement[9] in closed systems.  A stochastic evolution of the state vector can be arranged to yield a Lindblad equation for the density matrix, and with a  suitable choice of the details of this differential equation, its solutions can reproduce the results of measurement according to the Copenhagen interpretation, but through a smooth spontaneous localization of the density matrix[9] rather than a sudden intrusion of classical physics.  (These matters are discussed in a separate paper[10].)  These theories share the well-known feature of ordinary quantum mechanics, that entanglement does not lead to communication at a distance between isolated systems.  This is because (as explained below, in a more general context) nothing that is done in one system can instantaneously affect the density matrix of another system, though it can affect the state vector; also, all predictions can be derived from the density matrix, without knowing anything about state vectors, and the evolution of the density matrix depends only on the density matrix, not on the state vector.  But from the point of view explored in the present work, the study of the stochastic evolution of the state vector is unnecessary; it is only the differential equation for the density matrix that matters.

The time-translation symmetry transformations used in deriving the Lindblad equation take us only into the future, not the past, and hence form a semi-group, not a group.  If we are willing to consider new ways that the density matrix might transform under time-translation, then we ought to do the same for general symmetry groups, not just semi-groups.   This proposal runs into potential difficulties, each of which can be escaped through a narrow loophole.  As shown in Section II, in order to allow for any new group transformation rules, we would need to restrict the class of Hermitian operators that represent physical quantities.  For continuous symmetries, it would also be necessary to restrict the class of physically realizable density matrices, as described in Section VII.  Finally, in general it would also be necessary to suppose that the usual arguments for complete positivity do not apply to real physical systems, for reasons given in Section VIII.  If further study shows that these loopholes are not actually open, then on the basis of the arguments of this paper, we could conclude that, even in a quantum mechanics without state vectors, the density matrix transforms under symmetry groups just as in ordinary quantum mechanics.

\vspace{20pt}

\begin{center}
{\bf II. PHYSICAL QUANTITIES AND UNUSUAL SYMMETRIES}
\end{center}

To explore  unusual possibilities for symmetry transformations, we need first to say what we mean by a symmetry transformation.  
We will take a symmetry transformation to be a linear mapping $\rho\mapsto g(\rho)$ of density matrices,  which preserve their Hermiticity, positivity,  and unit trace.  For any such transformation $g$, we further assume that there is a corresponding linear transformation $A\mapsto g(A)$ of any operator $A$ representing a physical quantity, which preserves its Hermiticity,  such that the mean value of the physical quantity is left invariant:
\begin{equation}
{\rm Tr} \Big(g(A)\,g(\rho)\Big)={\rm Tr} \Big(A\,\rho\Big)\;.
\end{equation}

With this definition of symmetry transformations, it is important to decide just what operators can represent physical quantities.  Certainly we want to include familiar quantities like momentum, angular momentum, etc. and functions of these quantities.  In particular, the projection operator on a non-degenerate eigenstate of such a physical quantity, as for instance the projection operator $(1\pm2s_z)/2$ on a state of a spin $1/2$ with $s_z=\pm 1/2$, represents a physical quantity.   In ordinary quantum mechanics, {\em any} Hermitian operator is assumed to represent a physical quantity, but if state vectors are not to be taken as a representation of reality, we can doubt whether  operators $\Lambda_\Psi$  that are defined  as the projection operators on arbitrary state-vectors $\Psi$  necessarily represent   physical quantities.  If they did, then according to our interpretive postulate ${\rm Tr}(\Lambda_\Psi\,\rho)$ would be the probability that a system with density matrix $\rho$ is in a state $\Psi$, which is just the sort of statement that we are taking as generally meaningless.  In any case, it is hard to see how one could ever tell that Schr\"{o}dinger's cat is in a state $|alive>+|dead>$ rather than, say, $|alive>-|dead>$.  

This point is important for us,  because  if all Hermitian operators including projection operators represent physical quantities, then with our assumptions it can be shown that density matrices transform under any symmetry transformation $g$ with an inverse just as in ordinary quantum mechanics[11]:
\begin{equation}
 g(\rho)=U\,\rho\,U^\dagger\;,
\end{equation}
with $U$ unitary or antiunitary.

The first step in the proof is to show that if all projection operators are physical quantities in the above sense, then any symmetry transformation $g$ with an inverse $g^{-1}$ takes any   projection operator $\Lambda$ (defined here as a Hermitian operator with $\Lambda^2=\Lambda$ and ${\rm Tr}\Lambda=1$)  into another projection operator.    According to our definition of symmetries,  any density matrix $\rho$ is mapped into a Hermitian positive matrix  $g(\rho)$ with unit trace, which can therefore be expressed as
$$g(\rho)=\sum_n P_n\Lambda_n\;,$$
where $\Lambda_n$ are projection operators, satisfying $\Lambda_n\Lambda_m=\delta_{nm}\Lambda_n$ and ${\rm Tr}\Lambda_n=1$, and the $P_n$ are positive real numbers with $\sum_n P_n=1$.  We then have
$$\rho=\sum_n P_n\,g^{-1}(\Lambda_n)\;.$$
If all projection operators represent physical quantities, then we can use Eq.~(1) with any $\Lambda_n$ in place of $A$, $\rho$ taken as any $\Lambda_m$, and $g$ replaced with $g^{-1}$, so that
$${\rm Tr}\Big(g^{-1}(\Lambda_n)\,g^{-1}(\Lambda_m)\Big)={\rm Tr}\Big(\Lambda_n\,\Lambda_m\Big)=\delta_{nm}\;.$$
Hence 
$${\rm Tr}\Big(\rho^2\Big)=\sum_n P_n^2\;.$$
Now, if $\rho$ is a projection operator, then $\rho^2=\rho$, so ${\rm Tr}\Big(\rho^2\Big)={\rm Tr}\Big(\rho\Big)=1$,
and therefore $\sum_n P_n^2=1$.  But the only way that this can be satisfied by a set of real positive numbers $P_n$ with $\sum_n P_n=1$ is to have all $P_n$ vanish except for one, say $P_1$, with the value $P_1=1$.  Then $g(\rho)$ is itself a projection operator, namely $\Lambda_1$.

The rest of the proof is completed quickly. Any projection operator $\Lambda$ can be expressed as a dyad, $\Lambda_\Psi=\Psi\Psi^\dagger$, where $\Psi$ is a normalized state vector, unique up to a phase.  (This is the exception mentioned in Section I to the rule that density   matrices may be expressed in various different ways as linear combinations  of projection operators; the only such representation of a projection operator is as a unique dyad.)  Since as we have seen any symmetry transformation $g$ with an inverse takes projection operators into projection operators, we must have $g(\Lambda_\Psi)=\Lambda_{g(\Psi)}$, where the state vector $g(\Psi)$ is unique up to a phase.  Again, if these projection operators are to be taken as representing physical quantities, then in Eq.~(1) we can take $A=\Lambda_\Psi$ and $\rho=\Lambda_\Phi$ for any two state vectors $\Psi$ and $\Phi$, and find 
$${\rm Tr}\Big(\Lambda_{g(\Psi)}\,\Lambda_{g(\Phi)}\Big)={\rm Tr}\Big(\Lambda_\Psi\,\Lambda_\Phi\Big)$$
 and therefore
$$\left|\Big(g(\Psi),g(\Phi)\Big)\right|^2=\left|\Big(\Psi,\Phi\Big)\right|^2\;.$$
According to Wigner's theorem[12], if this condition is satisfied for all normalized state vectors $\Phi$ and $\Psi$, then it must be possible to choose the phases of all $g(\Psi)$  so that,
$$ g(\Psi)=U_g\Psi$$
where $U_g$ is a unitary (or antiunitary) operator, the same for all $\Psi$.  In this case, we have $g(\Lambda_\Psi)=U_g \Lambda_\Psi U_g^\dagger$.  Since any density matrix can be expressed (though not uniquely) as a linear combination of projection operators, they also transform as in Eq.~(2), as was to be proved.

We thus have a choice.  We can assume that the invariance condition (1) holds for all density matrices $\rho$ and all Hermitian operators $A$, in which case density matrices can only have the same transformation properties (2) as in ordinary quantum mechanics.  Or we can limit the validity of Eq.~(1) to a class of physical quantities that does not include projection operators on every state vector, in which case density matrices may have a much wider variety of symmetry transformation properties.  In this paper we will explore the consequences of the latter choice.

The behavior  of the density matrix under general symmetry transformations is outlined here in Section III.  In Section IV these general results are applied and further explored for the case of continuous symmetries.  The group multiplication law is found to impose severe constraints on the transformation of the density matrix.  Section V presents an example of a class of  continuous symmetries whose action on the density matrix explicitly satisfies these constraints, but is different from the transformation found in ordinary quantum mechanics.  Section VI describes special features of the action of compact groups on the density matrix.  Section VII takes up the important but difficult question of deciding what conditions should be imposed on the transformation of the density matrix under general symmetry operations so that these transformations will preserve the positivity of the density matrix.  Section VIII shows that assuming the positivity of the eigenvalues of the transformation kernel rules out the possibility that the density matrix transforms differently than in ordinary quantum mechanics, and gives reasons why these eigenvectors need not be positive.

\vspace{20pt}

\begin{center}
{\bf III. GENERAL SYMMETRIES}
\end{center}

We suppose that a general element $g$ of the symmetry group of a system induces on the density matrix a linear transformation $\rho\mapsto g(\rho)$, with:
\begin{equation}
g(\rho)_{M'N'}=\sum_{MN}K_{M'M,N'N}[g]\,\rho_{MN}\;,
\end{equation}
where $K[g]$ is some c-number kernel independent of $\rho$.  
We will take the indices $M$, $N$, etc. to run here over a finite number $d$ of values, but will assume that the formalism can be extended to Hilbert spaces of infinite dimensionality, on which the matrices considered here become well-behaved operators.   (No attempt will be made here to apply this formalism to relativistic  theories[13].)  Our reason for concentrating on linear transformations is explained later in this section.

In order for $g(\rho)$ to be Hermitian for an arbitrary Hermitian $\rho$, it is necessary and sufficient that $K$ be Hermitian, in the sense that
 \begin{equation}
K_{M'M,N'N}[g]^*=K_{N'N,M'M}[g]\;.
\end{equation}
(This is why it proves convenient to put the subscripts on $K[g]$ in what may otherwise look like a peculiar order.)
Also, in order for $g(\rho)$ to have unit trace for an arbitrary $\rho$ with unit trace, it is necessary and sufficient that
 \begin{equation}
\sum_{M'}K_{M'M,M'N}[g]=\delta_{MN}\;.
\end{equation}
The difficult thing is to know  what additional conditions should be imposed on $K[g]$ (or on $\rho$) so that  $g(\rho)$ will be positive.  This will be discussed in Sections VII and VIII.

The great physical advantage of basing quantum mechanics on the density matrix, with linear symmetry transformation properties, is that the transformation properties of the density matrix for an isolated subsystem do not depend on the properties of any other distant isolated subsystem, even in the case of entanglement, where the density matrix does not factorize into density matrices for the individual subsystems.  Suppose that the system consists of two isolated parts, subsystems {\em I} and {\em II}, and replace the indices $M$, $N$, etc. with compound indices $ma$, $nb$, etc.,  with the first letter labeling  the states of subsystem {\em I} and the second the states of subsystem {\em II}.  The possibility of entanglement does not in general allow the density matrix to factor into a product $\rho^{(I)}_{mn}\rho^{(II)}_{ab}$ of density matrices for the two subsystems, but if the subsystems are isolated they transform independently, in the sense that the kernel in Eq.~(3) does factorize:
\begin{equation}
K_{m'a'ma,n'b'nb}[g]=K^{(I)}_{m'm,n'n}[g]\;K^{(II)}_{a'a,b'b}[g]\;,
\end{equation}
where $K^{(I)}[g]$ and $K^{(II)}[g]$ are the kernels that would describe the transformation of the density matrix in subsystems {\em I} and {\em II} if the other subsystem did not exist.  
(For a nonlinear transformation it would be  difficult to see what could take the place of Eq.~(6) as a statement of   what we mean by {\em isolated} subsystems.)
Since both $K^{(I)}[g]$ and $K^{(II)}[g]$ are possible physical kernels, they each satisfy the analog of Eq.~(5):
\begin{equation}
\sum_{m'}K^{(I)}_{m'm,m'n}[g]=\delta_{mn}\;,~~~\sum_{a'} K^{(II)}_{a'a,a'b}[g] =\delta_{ab}\;.
\end{equation}
  The density matrix of subsystem {\em I} is related to the density matrix of the whole system by
\begin{equation}
\rho_{mn}^{(I)}=\sum_a\rho_{ma,na}\;.
\end{equation}
(This follows from the requirement that the mean value ${\rm Tr}(\rho A)$ of any physical quantity represented by an operator of the form $A_{ma,nb}=A^{(I)}_{mn}\delta_{ab}$, which acts non-trivially only on subsystem {\em I}, should be equal to ${\rm Tr}(\rho^{(I)} A^{(I)})$.)
According to Eqs.~(3), (6) and (8), its transformation is given by
$$
\rho_{m'n'}^{(I)}\mapsto g^{(I)}(\rho)_{m'n'}=\sum_{a'}\sum_{mnab}\;K^{(I)}_{m'm,n'n}[g]\;K^{(II)}_{a'a,a'b}[g]\;\rho_{ma,nb}\;.
$$
Using Eq.~(7) for $K^{(II)}$ and Eq.~(8) again, this is
\begin{equation}
g^{(I)}(\rho)_{m'n'}=\sum_{mn}K^{(I)}_{m'm,n'n}[g]\;\rho_{mn}^{(I)}\;,
\end{equation}
so the transformation of $\rho^{(I)}$ is independent of $\rho^{(II)}$.  
This applies in particular to the symmetry of time-translation, so as well known even in entangled states the evolution of the density matrix for subsystem {\em I} is unaffected by whatever happens in subsystem {\em II}.

Now let us return to the general case, and our former notation.  Because the kernel $K[g]$ is Hermitian in the sense of Eq.~(4), it can be expanded as
\begin{equation}
K_{M'M,N'N}[g]=\sum_i \eta^{(i)}[g]\;u^{(i)}_{M'M}[g]\;u^{(i)*}_{N'N}[g]\;,
\end{equation}
where the $u^{(i)}_{M'M}[g]$ and $\eta^{(i)}[g]$ are a complete set of normalized eigenmatrices and eigenvalues of the kernel $K_{M'M,N'N}[g]$, in the sense that
\begin{equation}
\sum_{N'N}K_{M'M,N'N}[g]\,u^{(i)}_{N'N}[g]=\eta^{(i)}[g]\,u^{(i)}_{M'M}[g]\;,
\end{equation}
\begin{equation}
{\rm Tr}\Big(u^{(i)\dagger}[g]\,u^{(j)}[g]\Big)=\delta_{ij}\;.
\end{equation}
(Note that Eq.~(11) does {\em not} say that the map (3)  takes $u^{(i)}[g]$ into  $\eta^{(i)}[g]\,u^{(i)}[g]$.)
The transformed density matrix (3) can then be written more compactly as a sum of  matrix products:
\begin{equation}
g(\rho)=\sum_i \eta^{(i)}[g]\;u^{(i)}[g]\;\rho\;u^{(i)\dagger}[g]\;.
\end{equation}
The  trace condition (5) here reads
\begin{equation}
\sum_i \eta^{(i)}[g]\;u^{(i)\dagger}[g]\;u^{(i)}[g]=\BM{1}\;,
\end{equation}
where $\BM{1}$ is the unit matrix.
If there were only one eigenvector $u^{(1)}[g]$ with a non-zero eigenvalue $\eta^{(1)}[g]$, then Eq.~(14) would require $\eta^{(1)}[g]>0$, and the transformation rule (13) could be written $g(\rho)=U[g]\;\rho\;U^{\dagger}[g]$, where according to Eq.~(14) the matrix $U[g]\equiv \sqrt{\eta^{(1)}[g]}u^{(1)}[g]$ appearing in this transformation rule  is unitary.    But in the general case, where the kernel has  several independent eigenmatrices with non-zero eigenvalues,  Eqs.~(13) and (14)  represent  a non-trivial generalization of the unitary transformations of ordinary quantum mechanics.

We also need to impose on $K[g]$ the group property, that for any two symmetry transformations $g$ and $\overline{g}$, we have
\begin{equation}
\sum_{M'N'}K_{M''M',N''N'}[g]\;K_{M'M,N'N}[\overline{g}]=K_{M''M,N''N}[g\overline{g}]\;.
\end{equation}
(This condition is not usually mentioned in connection with time translation, because as we shall see it does not constrain the differential equation that governs the temporal evolution of the density matrix, but it does need to be imposed even for time translation when that symmetry is combined with other symmetries.)
Using the representation (10), the group property (15) may be written
\begin{eqnarray}
&&\sum_{M'N'}\sum_{ij}\eta^{(i)}[g]\,\eta^{(j)}[\overline{g}]\,u^{(i)}_{M''M'}[g]\,u^{(j)}_{M'M}[\overline{g}]\,u^{(i)\dagger}_{N'N''}[g]\,u^{(j)\dagger}_{NN'}[\overline{g}]\nonumber\\&&
~~~~~~~~~~=\sum_k \eta^{(k)}[g\overline{g}]\,u^{(k)}_{M''M}[g\overline{g}]\,u^{(k)\dagger}_{NN''}[g\overline{g}]\;,
\end{eqnarray}
or, in an abbreviated notation,
\begin{equation}
\sum_{ij}\eta^{(i)}[g]\,\eta^{(j)}[\overline{g}]\,u^{(i)}[g]\,u^{(j)}[\overline{g}]\;\otimes\; u^{(j)\dagger}[\overline{g}]\,u^{(i)\dagger}[g]
=\sum_k \eta^{(k)}[g\overline{g}]\,u^{(k)}[g\overline{g}]\otimes u^{(k)\dagger}[g\overline{g}]\;,
\end{equation}
it being understood that for any two $d\times d$ matrices $A$ and $B$, 
\begin{equation}
[A\otimes B]_{M'M,N'N}\equiv A_{M'M}\;B_{NN'}\;.
\end{equation}
In the next section we will explore the implications of Eq.~(17) for continuous symmetries.

\vspace{20pt}

\begin{center}
{\bf IV. CONTINUOUS SYMMETRIES}
\end{center}

We now consider a group of transformations that includes elements arbitrarily close to the identity ${\bf I}$.    For the identity, we have of course
\begin{equation}
K_{M'M,N'N}[{\bf I}]=\delta_{M'M}\delta_{N'N}\;.
\end{equation}
This has one eigenmatrix $u^{(1)}[{\bf I}]$ with non-zero eigenvalue
\begin{equation}
u^{(1)}_{N'N}[{\bf I}]=\delta_{N'N}/\sqrt{d}~~~~~~~~~~\eta^{(1)}[{\bf I}]=d\;,
\end{equation}
and $ d^2-1$ eigenmatrices $u^{(\alpha)}[{\bf I}]$, a complete set of traceless matrices, all  with  eigenvalues zero:
\begin{equation}
{\rm Tr}\; u^{(\alpha)}[{\bf I}]=0~~~~~~~~~~\eta^{(\alpha)}[{\bf I}]=0\;.
\end{equation}

Now let's consider  group elements $g(\epsilon n)$, with  $g(0)={\bf I}$, where $\epsilon$ is infinitesimal, and $n^r$ is a real vector specifying a fixed direction in the space of group parameters near the origin. 
  The kernel $K[g(\epsilon n)]$ may be supposed to be analytic in $\epsilon n$ for $\epsilon n$ near zero, but because the eigenvalues $\eta^{(\alpha)}[g(\epsilon n)]$ are degenerate for $\epsilon=0$, according to the familiar rules of first order perturbation theory the corresponding unperturbed eigenmatrices  must be chosen to diagonalize the first-order perturbation to $K[g(\epsilon n)]$, and  therefore may remain functions of the direction (but not of the magnitude) of $n$ even  for $\epsilon\rightarrow 0$.  That is, in the limit $\epsilon\rightarrow 0$, the   $u^{(\alpha)}[g(\epsilon n)]$ approach $u^{(\alpha)}(n)$, where:
\begin{equation}
\sum_{M'M N'N} u^{(\alpha)*}_{M'M}(n)\left[\frac{\partial K_{M'M ,N'N}[g(\epsilon n)]}{\partial \epsilon}\right]_{\epsilon=0}u^{(\beta)}_{N'N}(n)=\delta_{\alpha\beta}\,\Delta^{(\alpha)}(n)\,\;,
\end{equation}
where $\Delta^{(\alpha)}(n)$   scales as $\Delta^{(\alpha)}(c n)=c\Delta^{(\alpha)}(n)$, but like $u^{(\alpha)}(n)$ is not in general analytic in $n$ at $n=0$.  To first order in $\epsilon$, the corresponding eigenvalues are
\begin{equation}
\eta^{(\alpha)}[g(\epsilon n)]\rightarrow \epsilon\; \Delta^{(\alpha)}(n)\;.
\end{equation}

On the other hand, the eigenvalue $\eta^{(1)}[g(\epsilon n)]$ is not degenerate and does not vanish for $\epsilon=0$, so the quantity $\sqrt{\eta^{(1)}[g(\epsilon n)]}u^{(1)}[g(\epsilon n)]$, which appears in the terms in Eq.~(10) with $i$ or $j$ or $k$ equal to 1, may be supposed to be given by a power series in $\epsilon\,n$:
\begin{equation}
\sqrt{\eta^{(1)}[g(\epsilon n)]}u^{(1)}[g(\epsilon n)]\rightarrow 1-i\epsilon \;n\cdot \tau+O(\epsilon^2)\;,
\end{equation}
with $n\cdot \tau\equiv \sum_r n^r\tau_r$, where $[\tau_r]_{N'N}$  are constant matrices (not necessarily Hermitian),  independent of $\epsilon$ and $n$.  

The trace condition (14) tells us the anti-Hermitian parts of the matrices $\tau_r$:
\begin{equation}
-in\cdot \tau  +i n\cdot\tau^\dagger+\sum_\alpha \Delta^{(\alpha)}(n)\,u^{(\alpha)\dagger}(n) u^{(\alpha)}(n)=0\;.
\end{equation}
Now, $\Delta^{(\alpha)}(n)$ and  $u^{(\alpha)\dagger}(n)$ are complicated functions of the vector $n$ that defines a infinitesimal group element, but we can easily see that the sum in Eq.~(25) is simply linear in the components of $n$.   By using Eqs.~(10), (23), (24), and (25), we can calculate the kernel for this group element to first order in $\epsilon $. In the notation of Eq.~(18),
\begin{equation}
K[g(\epsilon n)]\rightarrow {\BM 1}\otimes {\BM 1}+\epsilon\left[-i\, n\cdot \tau\otimes \BM{1}+\BM{1}\otimes \,in\cdot\tau^\dagger
+\sum_\alpha \Delta^{(\alpha)}(n)\,u^{\alpha}(n)\otimes u^{\alpha\dagger}(n)\right]\;.
\end{equation}
We are assuming that the kernel itself is analytic in the group parameters near the origin, so that 
the first-order term in $K[g(\epsilon n)]$ must be of the form $\epsilon 
\sum_r n^r K_r$, with $K_r$ independent of $n$, and therefore also
\begin{equation}
\sum_\alpha \Delta^{(\alpha)}(n)\,u^{(\alpha)}(n)\otimes u^{(\alpha)\dagger}(n)=\sum_r n^r L_r
\end{equation}
with $L_r$ independent of $n$.  As a corollary, if we use Eq.~(18) to put indices back in Eq.~(27), and contract the indices $N'$ and $M'$, we learn that 
\begin{equation}
\sum_\alpha \Delta^{(\alpha)}(n)\,u^{(\alpha)\dagger}(n) u^{(\alpha)}(n)=\sum_r n^r \theta_r\,
\end{equation}
where the  $d\times d$ Hermitian matrix $\theta_r$ is given by $[\theta_r]_{NM}=\sum_{M'}[L_r]_{M'M,M'N}$, and is therefore independent of $n$.
We may therefore write
\begin{equation}
\tau_r=T_r-\frac{i}{2}\theta_r\;,
\end{equation}
where $T_r$ are  Hermitian matrices, which like $\tau_r$ and $\theta_r$ are independent of $n$.  

Using these results in Eq.~(13), we find the first-order change in the density matrix due to the transformation $g[\epsilon n]$:
\begin{eqnarray}
&&\delta_{\epsilon n} \rho=i\epsilon\,[n\cdot T,\rho]+\epsilon\sum_\alpha  \Delta^{(\alpha)}(n)\,\Bigg[u^{(\alpha)}(n)\,\rho\, u^{(\alpha)\dagger}(n)\nonumber\\&&~~~
-\frac{1}{2}u^{(\alpha)\dagger}(n)u^{(\alpha)}(n)\,\rho
-\frac{1}{2}\rho\, u^{(\alpha)\dagger}(n)u^{(\alpha)}(n)\Bigg]\;.
\end{eqnarray}
It is the set of matrices $T_r$ that here play a role like the usual Hermitian  matrix representation of the Lie algebra, though as we shall see it is only in special cases that they can be shown to satisfy the same commutation relations.

It may be noted that in itself the transformation rule (30) does not uniquely fix the matrices $T_r$ and $u^{(\alpha)}(n)$.  Without changing $\delta_{\epsilon n}\rho$, we may shift these matrices by
\begin{eqnarray*}
&& \Delta \,u^{(\alpha)}(n)={\BM 1}\,{\rm Tr}\left(C u^{(\alpha)}(n)\right)\;,\\
&& [\Delta \,T_r]_{M'M}=\frac{i}{2}\sum_{N'N} [L_r]_{M'M,NN'}C_{N'N}-\frac{i}{2}\sum_{N'N} [L_r]^*_{MM',NN'}C^*_{N'N}\;,
\end{eqnarray*}
where $C$ is an arbitrary complex matrix.  This allows us to make the trace of $u^{(\alpha)}(n)$ anything we like, for if we take $C=\sum c_\beta u^{(\beta)\dagger}(n)$ with $c_\beta$ arbitrary, then (using Eq.~(12)) we have ${\rm Tr}[\Delta u^{(\alpha)}(n)]=c_\alpha d$.  However in this paper we will stick to the original definitions of 
$T_r$ and $u^{(\alpha)}(n)$, characterized by the tracelessness of $u^{(\alpha)}(n)$.

So far,  this section has closely followed the usual treatment of the symmetry of time-translation, especially as in  ref. [14].  This symmetry  yields the Lindblad  equation[8] for the time dependence of the density matrix (given here in Section VIII), which applies in some extended versions of quantum mechanics[9].  In that case there is just one matrix  $T$, which can be identified with minus the Hamiltonian of the system.

We will now see what can be learned from the  multiplication rule (17) for general continuous groups, when $g=g(\epsilon n)$ and $\overline{g}=g(\epsilon\overline{n})$ are both near the identity.  Eq.~(17) is automatically satisfied if either $g={\bf I}$ or $\overline{g}={\bf I}$, so the lowest-order non-trivial terms in Eq.~(17) are of order $\epsilon^2$.  The resulting condition is a terrible mess, involving many coefficients that only reflect how group elements are parameterized.   To focus only on physically interesting quantities, we will ignore everything but the part of Eq.~(17) that is antisymmetric in $n$ and $\overline{n}$, which must be satisfied separately from the rest.   Eq.~(17) would be symmetric in $n$ and $\overline{n}$ if it were not for the non-vanishing commutators of the matrices $u^{(i)}$ on the left side of the equation and of the group elements themselves on the right side.  To calculate the latter terms, we may write                  
\begin{equation}
g(\epsilon n)g(\epsilon \overline{n})=g\Big(\epsilon n+\epsilon\overline{n}+\epsilon^2\,f(n,\overline{n})+\dots\Big)~~~~f^r=\frac{1}{2}\sum_{st}C^r_{st}n^s\overline{n}^t\;,
\end{equation}
where $C^r_{st}=-C^r_{ts}$ are the structure constants of the group's Lie algebra, and the dots in Eq.~(31) denote second-order terms that are symmetric in $n$ and $\overline{n}$, as well as terms of higher order in $\epsilon$.  The antisymmetric part of the terms in Eq.~(17) of order $\epsilon^2$ now gives
\begin{eqnarray}
&& [n\cdot\tau,\overline{n}\cdot\tau]\otimes {\BM 1}+{\BM 1}\otimes[n\cdot\tau,\overline{n}\cdot\tau]^\dagger\nonumber\\&&+i\sum_\alpha \Delta^{(\alpha)}(\overline{n})\;[\tau\cdot n,u^{(\alpha)}(\overline{n})]\otimes u^{(\alpha)\dagger}(\overline{n})-i\sum_\alpha \Delta^{(\alpha)}(\overline{n})\;
u^{(\alpha)}(\overline{n})\otimes [\tau\cdot n,u^{(\alpha)}(\overline{n})]^\dagger\nonumber\\&&
-i\sum_\alpha \Delta^{(\alpha)}(n)\;[\tau\cdot \overline{n},u^{(\alpha)}(n)]\otimes u^{(\alpha)\dagger}(n)+i\sum_\alpha \Delta^{(\alpha)}(n)\;
u^{(\alpha)}(n)\otimes [\tau\cdot \overline{n},u^{(\alpha)}(n)]^\dagger\nonumber\\&&
-\frac{1}{2}\sum_{\alpha\beta}\Delta^{(\alpha)}(n)\,\Delta^{(\beta)}(\overline{n})\;[u^{(\alpha)}(n),u^{(\beta)}(\overline{n})]\otimes \{u^{(\alpha)}(n),u^{(\beta)}(\overline{n})\}^\dagger\nonumber\\&&~~~
-\frac{1}{2}\sum_{\alpha\beta}\Delta^{(\alpha)}(n)\;\Delta^{(\beta)}(\overline{n})\;\{u^{(\alpha)}(n),u^{(\beta)}(\overline{n})\}\otimes [u^{(\alpha)}(n),u^{(\beta)}(\overline{n})]^\dagger\nonumber\\&&
=i\sum_{rst}\tau_r\,C^r_{st}n^s\overline{n}^t\otimes {\BM 1}-i{\BM 1}\otimes\sum_{rst}\tau^\dagger_r\,C^r_{st}n^s\overline{n}^t\nonumber\\&& -\sum_{rst} \left[\frac{\partial}{\partial (n+\overline{n})^r}\sum_\alpha\Delta^{(\alpha)}(n+\overline{n})u^{(\alpha)}(n+\overline{n})\otimes u^{(\alpha)\dagger}(n+\overline{n})\right]_{n+\overline{n}=0}\,C^r_{st}n^s\overline{n}^t
\end{eqnarray}
where curly brackets denote an anticommutator.

Inspection of Eq.~(30) shows that if all $\Delta^{(\alpha)}$ vanish then $\delta_{\epsilon n}\rho=i\epsilon[n\cdot T, \rho]$.  Further, Eqs.~(28) and (29) show in this case that  $\tau_r=T_r$.   Eq.~(32) then shows also that in this case the Hermitian matrices $T_r$  satisfy the commutation relations $[T_s,T_t]=i\sum_r C^r_{st}T_t$ of the symmetry group's Lie algebra.  

But these familiar results do not hold if the density matrix transforms in a more general way, with non-vanishing values for some $\Delta^{(\alpha)}$, in which case the terms in Eq.~(30), (28), and (32) with non-zero $\Delta^{(\alpha)}$  represent a departure from ordinary quantum mechanics.  In ordinary quantum mechanics the structure of the Hamiltonian and other operators representing symmetry generators is largely fixed by the condition that they satisfy the Lie algebra of the symmetry group, as for instance the  form of the kinetic energy terms in the non-relativistic Hamiltonian is fixed by the commutators of the generator of time-translation with the other generators of the Galilean group.  In the generalization of quantum mechanics considered here, it is Eqs.~(32) and (29) that must be used to constrain the operators $T_r$ and  $u^{(\alpha)}(n)$ that define the transformation of the density matrix.

\vspace{20pt}

\begin{center}
{\bf V. AN EXAMPLE}
\end{center}

The condition (32)  sets constraints on the sorts of matrices $T_r$ and $u^{(\alpha)}(n)$ that can enter in the transformation (30) of the density matrix for a given set of structure constants $C^t_{rs}$.  This section will  give an explicit example showing how these constraints can be satisfied, in a way different from that of ordinary quantum mechanics.  

We will consider a group containing (perhaps among other things) two commuting symmetry operations, characterized by vectors $n^r$ and $\overline{n}^r$ in the space of group parameters, for which $\sum_{rs}n^r\overline{n}^sC_{rs}^t=0$.  (One of these symmetry operations may be time translation.)  To satisfy the constraints, let us try the assumption that the matrices $n\cdot T$, $\overline{n}\cdot T$, and the relevant $u^{(\alpha)}(n)$, $u^{(\alpha)\dagger}(n)$, $u^{(\beta)}(\overline{n})$, and $u^{(\beta)\dagger}(\overline{n})$ all commute with one another (where by relevant we mean that $\Delta^{(\alpha)}(n)$ and $\Delta^{(\beta)}(\overline{n})$ are not all zero.)     Then the definition (28), (29) shows that the matrices $n\cdot \tau$ and $\overline{n}\cdot \tau$ commute with each other and with the relevant $u^{(\alpha)}(n)$, $u^{(\alpha)\dagger}(n)$, $u^{(\beta)}(\overline{n})$, and $u^{(\beta)\dagger}(\overline{n})$.    The constraint (32) is then satisfied, as every term in this constraint simply vanishes.

As simple as this example is, it represents a non-trivial generalization of ordinary quantum mechanics.  Since the Hermitian $n\cdot T$ and the relevant $u^{(\alpha)}(n)$ and $u^{(\alpha)\dagger}(n)$ all commute with one another, we can choose a basis in which they are all diagonal, with
$$ [u^{(\alpha)}(n)]_{MN}=\delta_{MN}\;u_{\alpha M}(n)\;,~~~[n\cdot T]_{MN}=\delta_{MN}\;n\cdot T_{M}\;.$$
The transformation (30) then reads
\begin{eqnarray*}
&&\delta_{\epsilon n}\rho_{MN}=\epsilon\Bigg[in\cdot(T_{M}-T_{N})+\sum_\alpha \Delta^{(\alpha)}(n)\Big[u_{\alpha M}(n)\,u_{\alpha N}(n)^*
\\&&~~~~~~-\frac{1}{2}|u_{\alpha M}(n)|^2-\frac{1}{2}|u_{\alpha N}(n)|^2\Big]\Bigg]\rho_{MN}\;,
\end{eqnarray*}
so that $\rho_{MN}$ is not simply multiplied by a difference $f(M)-f(N)$ for some function $f$, as in ordinary quantum mechanics.  In the absence of any other symmetries the parameters $\Delta^{(\alpha)}(n)$, $n\cdot T_{M}$, and $u_{\alpha M}(n)$ would be  unconstrained, except that $\Delta^{(\alpha)}(n)$ and $T_{
M}$ are real.

\vspace{20pt}

\begin{center}
{\bf VI. COMPACT GROUPS}
\end{center}

A well-known theorem tells us that with a suitable choice of basis, all finite-dimensional representations of compact groups are unitary.  The density matrix furnishes a $d^2$-dimensional representation of any symmetry group, so  for compact groups it should transform unitarily.  As we will now see, this does {\em not} mean that  it undergoes the transformation $\rho\mapsto U\rho U^\dagger$ of ordinary quantum mechanics, but it does constrain its transformation properties in interesting ways, one of which will be important in dealing with the issue of positivity.

The unitarity of the transformation (3) requires that
\begin{equation}
\sum_{M''N''} K_{MM'',NN''}[g]\,K^*_{M'M'',N'N''}[g]=\sum_{M''N''} K^*_{M''M',N''N'}[g]\,K_{M''M,N''N}[g]=\delta_{M'M}\delta_{N'N}\;.
\end{equation}
One immediate consequence that we will need in Section VII is that the density matrix $\rho={\BM 1}/d$ is invariant.  We can see this by contracting the first equation (33) with $\delta_{MN}$ and using the trace condition Eq.~(5).  This gives
$$\sum_{M''}K^*_{M'M'',N'M''}[g]=\delta_{M'N'}\;.$$
Taking the complex conjugate and dividing by $d$ then gives the statement of invariance:
\begin{equation}
g({\BM 1}/d)={\BM 1}/d
\end{equation}

In the notation (18), Eq.~(33)  reads
\begin{eqnarray}
&&{\BM 1}\otimes {\BM 1}=\sum_{ij}\eta^{(i)}[g]\eta^{(j)}[g]u^{(i)}[g]u^{(j)\dagger}[g]\otimes u^{(j)}[g]u^{(i)\dagger}[g]\nonumber\\&&~~~~~~~
=\sum_{ij}\eta^{(i)}[g]\eta^{(j)}[g]u^{(i)\dagger}[g]u^{(j)}[g]\otimes u^{(j)\dagger}[g]u^{(i)}[g]\;.
\end{eqnarray}
For elements of continuous groups with parameters $\epsilon n^r$ near the origin, we can either use Eqs.~(24) and (29) in Eq.~(35), or use Eq.~(26) directly in Eq.~(33), and in either way find that 
\begin{equation}
n\cdot \theta\otimes {\BM 1}+{\BM 1}\otimes n\cdot \theta=\sum_\alpha \Delta^{(\alpha)}[n]\,u^{(\alpha)}(n)\otimes u^{(\alpha)\dagger}(n)
+\sum_\alpha \Delta^{(\alpha)}[n]\,u^{(\alpha)\dagger}(n)\otimes u^{(\alpha)}(n)\;.
\end{equation}
Taking the trace of the matrices on the right of the direct products then gives
$$
d\,n\cdot \theta+{\BM 1}\,{\rm Tr}(n\cdot \theta)=0\;.
$$
The trace of this equation gives $2d{\rm Tr}(n\cdot \theta)=0$, so $n\cdot \theta=0$, and therefore for compact groups 
\begin{equation}
\tau_r=T_r\;.
\end{equation}

\begin{center}
{\bf VII. POSITIVITY}
\end{center}

Finally, we come to the  issue of positivity.  It is clearly necessary that the linear mapping (3) corresponding to a symmetry transformation should take the density matrices of physical states into other density matrices that are positive as well as being Hermitian and having unit trace.  A linear mapping is itself called {\em positive} if takes {\em all} positive Hermitian matrices into positive Hermitian matrices with the same trace.  It would simplify matters if we could just assume that all symmetry mappings are positive, but this is not indispensable, because the density matrices of physical states may be limited in some way that insures that they are mapped into positive matrices, even if some other positive matrices are not mapped into positive matrices.  This is particularly plausible for compact symmetry groups, for which $g(\rho)$ for any $\rho$ varies only over a compact manifold.   For instance, in the $SU(3)$ example of Section I, the density matrix is positive if
(though not only if) it is subject to the inequality
$$ |b_1|^2+|b_2|^2+|b_3|^2\leq \frac{1}{4}a_1 a_2 a_3\;.$$
This condition is $SU(3)$-invariant, so under $SU(3)$ transformations any density matrix satisfying  this condition will be transformed into another density matrix satisfying the same condition, and will therefore  also be positive.

This is an important point, because we can show that if the mapping associated with  any invertible continuous symmetry  acts on all density matrices as a positive mapping, then this mapping must take the same form (2)  as in ordinary quantum mechanics.  For the purposes of this theorem, we only need to show that for any invertible mapping that is not of the form (2) there is  {\it some} positive density matrix $\rho$ that is transformed into a non-positive matrix, so we are free to choose $\rho$ here pretty much as we like.  We will choose the density matrix $\rho$ to have  one  eigenvector $v$ with eigenvalue zero:
\begin{equation}
\rho v=v^\dagger\rho=0\;.
\end{equation}
When we take the  expectation value of Eq.~(30) in the ``state'' $v$, as a consequence of Eq.~(38) we find that only  the first term in the sum over $\alpha$ makes a non-zero contribution:
\begin{equation}
\Big(v^\dagger \,[\rho+\delta_{\epsilon n}\rho]v\Big)= \Big(v^\dagger\, \delta_{\epsilon n}\rho\, v\,\Big)=\epsilon\sum_\alpha  \Delta^{(\alpha)}(n)\, \Big(v^\dagger u^{(\alpha)}(n)\,\rho\, u^{(\alpha)\dagger}(n)v\Big)\;.
\end{equation}
It is immediately obvious that if the coefficient of $\epsilon$ is non-zero, then for some sign of $\epsilon$ the expectation value (39) will be negative, so that
$\rho+\delta_{\epsilon n}\rho$ cannot be positive.  

It only remains to show that unless all $\Delta^{(\alpha)}(n)$ vanish, for some  vector $v$ there will be some positive Hermitian matrix $\rho$ satisfying Eq.~(38) for which the coefficient of $\epsilon$ in Eq.~(39) is non-zero.  (This is obvious if all $\Delta^{(\alpha)}(n)$ have the same sign, but we want also to consider the case where some are positive and some are negative.)  Let us suppose the contrary; that is, for all $v$ we have 
\begin{equation}
\sum_\alpha  \Delta^{(\alpha)}(n)\,\Big( v^\dagger u^{(\alpha)}(n)\,\rho\, u^{(\alpha)\dagger}(n)v\Big) =0
\end{equation}
for all positive Hermitian matrices $\rho$ satisfying Eq.~(38).  We will show that in this case, we must have $\Delta^{(\alpha)}(n)=0$ for all $\alpha$.  

We are free to take $\rho$ to have no eigenvectors with eigenvalue zero other than $v$. In this case, the condition that $\rho$ is positive puts no constraints on infinitesimal variations of $\rho$, so the assumption that Eq.~(40) holds for all positive Hermitian $\rho$ satisfying Eq.~(38) requires that 
\begin{equation}
\sum_\alpha  \Delta^{(\alpha)}(n)\, [v^\dagger u^{(\alpha)}(n)]_M\, [u^{(\alpha)\dagger}(n)v]_N =c_M^*v_N+v_M^*c_N\;,
\end{equation}
for all $N$ and $M$, and for some vector $c_N$ which may depend on $n$ and $v$.  In fact, $c_N$ must depend on $v$, because Eq.~(41) is supposed to hold for all $v$, so there have to be the same numbers of $v$s and $v^*$s on both sides of the equation.  Specifically, we must have $c_N=\sum_L C_{NL}v_L$, where $C_{NL}$ is independent of $v$, and the coefficient of $v_P^*v_Q$ in Eq.~(41) must vanish:
\begin{equation}
\sum_\alpha  \Delta^{(\alpha)}(n)\,  u_{PM}^{(\alpha)}(n)]\, u_{NQ}^{(\alpha)\dagger}(n) =C_{MP}^*\delta_{NQ}+\delta_{MP}C_{NQ}\;.
\end{equation}
Now we can use some of the properties of the $u^{(\alpha)}(n)$ obtained in Section IV.  Consider any $\beta$,  and contract Eq.~(42) with $u^{(\beta)^\dagger}_{MP}(n)u^{(\beta)}_{QN}(n)$.  Because the $u^{(\beta)}(n)$ are traceless, the right-hand side of the contracted equation   vanishes, and because they satisfy the orthonormality condition ${\rm Tr}\Big(u^{(\beta)^\dagger}u^{(\alpha)}\Big)=\delta_{\beta\alpha}$, the left-hand side of the  contracted equation is $\Delta^{(\beta)}(n)$, so $\Delta^{(\beta)}(n)=0$ for all $\beta$, as was to be shown.

This theorem leaves open the possibility of limiting the density matrix to a special class of positive Hermitian matrices, for which  any symmetry transformation takes any member of this class into another positive member of the same class, as in the $SU(3)$ example given above.  But does such a special class always exist?  Apparently it does, at least for compact groups.  We saw in the previous section that in a suitable basis, such symmetries leave invariant the positive density matrix ${\BM 1}/d$.  Suppose we shift this density matrix by some traceless Hermitian matrix $\eta$.  The new density matrix  ${\BM 1}/d+\eta$ will generally not  be invariant, but  as long as $\eta$ is sufficiently small the symmetry transformations  belonging to a compact group will take ${\BM 1}/d+\eta$ into  density matrices whose eigenvalues are sufficiently close to the original eigenvalues $1/d$ so that they are still all positive.  Thus,  acting  on a density matrix  ${\BM 1}/d+\eta$ with all symmetries $g$ belonging to a compact group, and with $\eta$  running over all traceless Hermitian matrices that are   sufficiently small so that all $g({\BM 1}/d+\eta)$  are positive, provides the sort of special class of density matrices we need, which can transform in a way that is different from the transformation (2) of   ordinary quantum mechanics without raising problems with positivity.

\vspace{20pt}

\begin{center}
{\bf VIII.  MAPPINGS WITH POSITIVE EIGENVALUES}
\end{center}

There is a class of mappings that are obviously positive, in the sense of taking all positive Hermitian matrices into positive Hermitian matrices.  Inspection of Eq.~(13)  shows immediately that a mapping is positive if (though not only if) all its eigenvalues $\eta^{(i)}$ are positive.  In this case, we can write the general mapping (13) in the Kraus form[15]:
\begin{equation}
g(\rho)=\sum_i A^{(i)}[g]\,\rho\,A^{(i)\dagger}[g]
\end{equation}
where $A^{(i)}[g]\equiv \eta^{(i)}[g]^{1/2} u^{(i)}[g]$ and $\sum_i A^{(i)\dagger}[g]A^{(i)}[g]=1$.  The  unitary transformations  $\rho\mapsto U[g]\,\rho\,U[g]^\dagger$ of ordinary quantum mechanics are a special case of the more general transformations (43). 

The assumption that all eigenvalues of the kernel are positive would be an effective way of guaranteeing that positive density matrices are mapped into positive density matrices, but it would have the  consequence that the transformation of the density matrix under any symmetry group, whether continuous or discrete, would reduce to the same unitary transformation rule (2)  as in ordinary quantum mechanics.  

We can see this immediately for continuous groups.  If for some continuous symmetry $g$ we were to require  the positivity of the eigenvalues (23) whatever the sign of $\epsilon$, we would have to assume that $\Delta^{(\alpha)}[g]=0$ for all $\alpha$.  As already mentioned in Section IV, any continuous symmetry for which $\Delta^{(\alpha)}[g]$ vanishes for all $\alpha$ is necessarily realized by the unitary transformation (2) of ordinary quantum mechanics.  Of course, we already knew this.  With all eigenvalues positive, mappings preserve the positivity of any density matrix, and as shown in the previous section any continuous symmetry with an inverse that preserves the positivity of all density matrices must act as in Eq.~(2).

As already mentioned in Section I,  in various extended versions of quantum mechanics[9] the assumption of positive mapping of the density matrix  is commonly made for the continuous symmetry of time-translation, but only prospectively, not retrospectively.  It is usually assumed in these theories that  the kernel  for time-translation by an amount $\tau$ has positive eigenvalues if $\tau>0$, but this is not assumed (and in fact is not generally true)  when $\tau<0$.  For this reason, the positive time-translation mappings usually considered in these theories form a  semi-group, not a group. 
With this weaker assumption, Eq,~(23) simply requires that $\Delta^{(\alpha)}(n_{\cal T})\geq 0$, where $n_{\cal T}$ denotes the direction in the space of  group parameters for time-translation.  The transformation rule (30) then immediately yields the Lindblad equation  
$$
\frac{d}{dt}\rho=-i[H,\rho]+\sum_\alpha\Bigg[L_\alpha\,\rho\, L_\alpha^\dagger-\frac{1}{2}L_\alpha^\dagger \,L_\alpha\,\rho-\frac{1}{2}\rho \,L_\alpha^\dagger\, L_\alpha\Bigg]\;,
$$
where $H\equiv -n_{\cal T}\cdot \cdot T$ and $L_\alpha\equiv \Delta^{(\alpha)1/2}[n_{\cal T}]\,u^{(\alpha)}[n_{\cal T}]$.
If $\rho(t)$ is positive for some initial time $t$ then this equation gives a positive $\rho(t')$ for all $t'>t$ but, as illustrated in  ref. [10], in these theories one can usually find an earlier time $t'<t$ for which $
\rho(t')$  is not positive.    As I understand it, this is tolerated in these extended versions of quantum mechanics because unless we tackle the description of the whole universe the differential equation for the time-dependence of the density matrix is only supposed to apply for closed systems, which become closed at some initial time $t$, so one does not have to worry about what happens for times $t'<t$.   
One can argue about whether this is satisfactory for time-translation, but we would certainly not want to assume that the eigenvalues of $K[g]$ are all positive when  the symmetry transformation $g$ is a spatial translation to the north but not to the south, or is a rotation that is clockwise around the vertical but not counter-clockwise, or is a boost that increases the velocity to the eastward but not to the westward.  Such symmetry transformations must be assumed to form a group, not merely a semi-group.

With a little more effort, we can show that for all symmetry groups, discrete as well as continuous, the assumption of positive eigenvalues rules out the possibility that the density matrix will have an unusual transformation rule, one different from Eq.~(2).   To prove this, we use one of the defining properties of groups, that for every element $g$ of a group there is an inverse $g^{-1}$.  From Eq.~(16) and Eqs.~(19)--(21), with $\overline{g}$ taken as $g^{-1}$, we have
\begin{equation}
\sum_{ij}\eta^{(i)}[g]\,\eta^{(j)}[g^{-1}]\,u^{(i)}[g]\,u^{(j)}[g^{-1}]\,\rho\,u^{(j)\dagger}[g^{-1}]\, u^{(i)\dagger}[g]=\rho\;,
\end{equation}
for any   matrix $\rho$.  In particular, for an Hermitian matrix $\rho$, we can find a unitary matrix $\Omega$ such that $\rho^{({\rm D})}=\Omega\rho\Omega^{-1}$ is  diagonal, $[\rho^{({\rm D})}]_{MN}=P_M\delta_{MN}$.  Eq.~(44) then applies if we replace $\rho$ with $\rho^{(D)}$ and replace all $u^{(i)}[g]$ and $u^{(j)}[g^{-1}]$ with $u^{(i{\rm D})}[g]\equiv \Omega u^{(i)}[g] \Omega^{-1}$ and $u^{(j{\rm D})}[g^{-1}]\equiv \Omega u^{(j)}[g^{-1}]\Omega^{-1}$.  This gives
\begin{eqnarray}
&&\sum_{ij}\eta^{(i)}[g]\,\eta^{(j)}[g^{-1}]\,\sum_L\Big[u^{(i{\rm D})}[g]\,u^{(j{\rm D})}[g^{-1}]\Big]_{ML}\,P_L\,\Big[u^{(i{\rm D})}[g]\,u^{(j{\rm D})}[g^{-1}] \Big]_{NL}^*\nonumber\\&&~~~~~~~~~~~=P_M\delta_{MN}\;,
\end{eqnarray}
This must hold for all real numbers $P_N$, so it follows that, for all $L$, $M$, and $N$:
\begin{equation}
\sum_{ij}\eta^{(i)}[g]\,\eta^{(j)}[g^{-1}]\,\Big[u^{(i{\rm D})}[g]\,u^{(j{\rm D})}[g^{-1}]\Big]_{ML}\,\Big[u^{(i{\rm D})}[g]\,u^{(j{\rm D})}[g^{-1}] \Big]_{NL}^*=\delta_{ML}\delta_{NL}\;.
\end{equation}
In particular, if $M=N\neq L$, then 
\begin{equation}
\sum_{ij}\eta^{(i)}[g]\,\eta^{(j)}[g^{-1}]\,\left|\Big[u^{(i{\rm D})}[g]\,u^{(j{\rm D})}[g^{-1}]\Big]_{ML}\right|^2=0\;.
\end{equation}
Here is where the positivity of the eigenvalues becomes important.  If all the eigenvalues 
$\eta^{(i)}[g]$ and $\eta^{(j)}[g^{-1}]$ are positive, then it follows from Eq.~(47) that for all relevant $i$ and $j$ (that is, for all $i$ and $j$ for which $\eta^{(i)}[g]$ and $\eta^{(j)}[g^{-1}]$ respectively do not vanish) we have
\begin{equation}
\Big[u^{(i{\rm D})}[g]\,u^{(j{\rm D})}[g^{-1}]\Big]_{ML}=0\;,
\end{equation}
for any unequal indices $M$ and $L$.
Since the matrix $u^{(i{\rm D})}[g]\,u^{(j{\rm D})}[g^{-1}]$ is thus diagonal, it commutes with 
the diagonal matrix $\rho^{(\rm D)}$.  But then also $u^{(i)}[g]\,u^{(j)}[g^{-1}]$ commutes with the arbitrary Hermitian matrix $\rho$.  The only matrices that commute with all Hermitian matrices are proportional to the unit matrix, so we can conclude that for all relevant $i$ and $j$
\begin{equation}
u^{(i)}[g]\,u^{(j)}[g^{-1}]=c_{ij}[g]{\BM 1}\;.
\end{equation}
for some complex numerical coefficients $c_{ij}[g]$.  Taking the determinant of Eq.~(49) gives $(c_{ij}[g])^d={\rm Det}  
u^{(i)}[g]\,{\rm Det}u^{(j)}[g^{-1}]$.  Now, there must be at least one relevant $j$ for which ${\rm Det}u^{(j)}[g^{-1}] \neq 0$, since otherwise we would have $u^{(i)}[g]\,u^{(j)}[g^{-1}]=0$ for all relevant $i$ and $j$, contradicting Eq.~(44).  Taking $j$ to have any value for which ${\rm Det}u^{(j)}[g^{-1}] \neq 0$, Eq.~(49) then tells us that all relevant $u^{(i)}[g]$ are proportional to a single $u[g]$; specifically,
\begin{eqnarray}
u^{(i)}[g]=\Big({\rm Det} u^{(i)}[g]\Big)^{1/d}\times u[g]\;,
\end{eqnarray}
where $u[g]=u^{(j)-1}[g^{-1}]\times {\rm \Big(Det}u^{(j)}[g^{-1}]\Big)^{1/d}$.    The trace condition (14) then reads 
$$ \sum_i \eta^{(i)}[g]\;\left |{\rm Det} u^{(i)}[g]\right|^{2/d} u^\dagger[g]u[g]=1\;,$$
so the matrix  $U[g]\equiv \left[\sum_i \eta^{(i)}[g]\;\left |{\rm Det} u^{(i)}[g]\right|^{2/d}\right]^{1/2} u[g]$ is unitary, and the transformation rule (13) takes  the familiar form $g(\rho)=U[g]\;\rho\;U^{\dagger}[g]$ of ordinary quantum mechanics, as was to be proved.

But do we need to require that the kernels for general symmetries have only positive eigenvalues?  There are well-known examples of positive mappings that have some negative eigenvalues.  The standard example is the transposition map
$$K_{M'M,N'N}=\delta_{M'N}\delta_{N'M}\;.$$
This has two eigenvalues, one positive and one negative.  (Any symmetric or antisymmetric matrix is an eigenmatrix with eigenvalue $+1$ or $-1$.)  Nevertheless,  $K$ is positive, because $g(\rho)=\rho^{\rm T}$, which is positive if $\rho$ is positive.

There is a widely cited argument for the requirement that all eigenvalues of the kernel $K$ must be positive,  based on the possibility of entanglement.   Consider an arbitrary system ${\cal S}^{(I)}$, and an arbitrary linear mapping $K^{(I)}$ of the density matrix of this system, which preserves its Hermiticity,  unit trace, and positivity.  We can imagine adding an isolated system ${\cal S}^{(II)}$ of finite dimensionality $d_{II}$, and extending $K^{(I)}$ to a kernel $K$ that acts as $K^{(I)}$ on ${\cal S}^{(I)}$, and  acts trivially on ${\cal S}^{(II)}$.  That is, if we label the basis vectors of ${\cal S}^{(I)}$ with indices $m$, $n$, etc. and the 
basis vectors of ${\cal S}^{(II)}$ with indices $a$, $b$, etc.,
 the kernel of the mapping (in the notation of Eqs.~(3) and (6)) on the combined system is 
\begin{equation}
K_{m'a'ma,n'b'nb}=K^{(I)}_{m'm,n'n}\delta_{a'a}\delta_{b'b}\;.
\end{equation}
The original mapping $K^{(I)}$ is said to be {\em completely positive}[16] if $K$ is positive (in the sense of mapping all positive density matrices for the combined system into positive density matrices) for all finite dimensionalities $d_{II}$.  A theorem due to Choi[17] states that if $K^{(I)}$ is completely positive in this sense, then all its eigenvalues are positive.  (As usually stated, the theorem says that any completely positive mapping takes the Kraus form (43), but as we have seen this form follows from the positivity of the eigenvalues, and it is obvious that any kernel that induces a transformation of this form has only positive eigenvalues.)

Though there is no doubt of the mathematical correctness of the Choi theorem,  it is not clear that it is relevant physically.  In particular, the vacuum is the only physical system that is invariant under Galilean (or Lorentz) transformations and time translations.  Since the dimensionality of the Hilbert space of the vacuum is unity, this does not fulfill the conditions of the Choi theorem, that there should be isolated systems ${\cal S}^{(II)}$ with arbitrary finite dimensionality on which the symmetry acts trivially.  

There seems to be a widespread impression that this does not matter, at least for the only symmetry that has been previously studied in this context, the symmetry of time translation.  It is supposed that, even if a symmetry transformation $K$ acts non-trivially on ${\cal S}^{(II)}$, we may be able to undo it by inventing a transformation $L$ that acts on ${\cal S}^{(II)}$ as the inverse of $K$, and leaves ${\cal S}^{(I)}$ unchanged, so that $LK$ does have the form (51).  (I have not been able to find a published reference to this argument.)   But in general, except in the uninteresting case in which ${\cal S}^{(I)}$ is the vacuum, $L$ will not be a symmetry transformation, so neither will be $LK$.  Or if we take $L$ as a symmetry transformation that acts non-trivially on ${\cal S}^I$, then the action of $LK$ on the Hilbert space of ${\cal S}^I$ is a completely positive mapping, but it is not the same mapping as  $K^{(I)}$.

There are some continuous symmetry transformations, such as rotations, for which there are invariant physical systems with Hilbert spaces of arbitrary dimensionality which  therefore might be taken as the isolated system ${\cal S}^{(II)}$ in the assumptions of the Choi theorem.  Even so, in the real world there are no disembodied spins, only particles with spins.  The Hilbert space of any physical system other than the vacuum has infinite dimensionality, and it is not clear that the Choi theorem can be extended to realistic cases.

Despite these skeptical comments, it may turn out to be physically necessary  for the kernels $K[g]$ for all elements $g$ of   symmetry transformation groups  to have only positive eigenvalues, even for symmetries like Galilean invariance.  In that case, the main point of  the of this paper would be the proof that such  symmetry transformations act on the density matrix as in ordinary quantum mechanics.  It would be much more interesting if for some symmetry groups it will turn out to be unnecessary for all eigenvalues to be positive, in which case the much richer variety of   symmetry transformations
discussed in Section II and III would be physically possible.  

\vspace{20pt}

\noindent
{\em Added note:}  This work was originally presented on February 28, 2014, at a conference in honor of Joseph Polchinski at the Kavli Institute for Theoretical Physics, Santa Barbara, CA.   I subsequently was sent two papers of J. A. Barandes and D. Kagan that also propose to regard the density matrix rather than state vectors as the representation of reality.  Distribution of the present paper has been  delayed in order to take account of comments of G. Moore on the talk given in Santa Barbara. 

\vspace{20pt}

Special thanks are due to Gregory Moore for his incisive comments on the original version of this work.  I am also grateful to Nicolas Gisin and Philip Pearle for helpful recent correspondence, and to Jacques Distler, Angelo Bassi, Gian Carlo Ghirardi, James Hartle, and Roderich Tumulka for interesting earlier communications on the interpretation of quantum mechanics.  This material is based upon work supported by the National Science Foundation under Grant Numbers PHY-1316033 and PHY-0969020 and with support from The Robert A. Welch Foundation, Grant No. F-0014.

\vspace{20pt}

\begin{center}
{\bf ---------}
\end{center}

\vspace{10pt}

\begin{enumerate}

\item N. Bohr, Nature {\bf 121}, 580 (1928).
\item The published version is H. Everett, Rev. Mod. Phys. {\bf 29}, 454 (1957).
 \item R. B. Griffiths, J. Stat. Phys. {\bf 36}, 219 (1984); R. Omn\`{e}s, Rev. Mod. Phys. {\bf 64}, 339 (1992); M. Gell--Mann and J. B. Hartle, in {\em Complexity, Entropy, and the Physics of Information}, ed. W. Zurek (Addison--Wesley, Reading, MA, 1990); in {\em Proceedings of the Third International Symposium on the Foundations of Quantum Mechanics in the Light of New Technology}, ed. S. Kobayashi, H. Ezawa, Y. Murayama, and S. Nomura (Physical Society of Japan, 1990); in {Proceedings of the 25th International Conference on High Energy Physics, Singapore, August 2--8, 1990}, ed. K. K. Phua and Y Yamaguchi (World Scientific, Singapore, 1990); J. B. Hartle, {\em Directions in Relativity, Vol. 1}, ed. B.-L. Hu, M.P. Ryan, and C.V. Vishveshwars (Cambridge University Press, Cambridge, 1993).  For a survey and more recent references, see P. Hohenberg, Rev. Mod. Phys. {\bf 82}, 2835 (2010).
  \item A. Einstein, B. Podolsky, and N. Rosen, Phys. Rev. {\bf 47}, 777 (1936); D. Bohm, {\em Quantum Theory} (Prentice-hall, Inc., New York, 1951), Chapter XXII; D. Bohm and Y. Aharonov, Phys. Rev. {\bf 108}, 1070 (1957).            
 \item  S. Weinberg, Phys. Rev. A {\bf 85}, 062116 (2012). 
\item N. Gisin, Helv. Phys. Acta {\bf 62}, 363 (1989); Phys. Lett. A {\bf 143}, 1 (1990).  
\item  J. Polchinski, Phys. Rev. Lett. {\bf 66}, 397 (1991).
\item G. Lindblad, Commun. Math. Phys. {\bf 48}, 119 (1976); V. Gorini, A. Kossakowski and E. C. G. Sudarshan, J. Math. Phys. {\bf 17}, 821 (1976).  The Lindblad equation can be derived as a straightforward application of an earlier result of A. Kossakowski, Reports on Math.  Phys. {\bf 3}, 247 (1972), Eq.~(77).
The equation was independently derived by T. Banks, L. Susskind, and M. E. Peskin, Nucl. Phys. B 244, 125 (1984).
\item G. C. Ghirardi, A. Rimini  
and T. Weber, Phys. Rev. D {\bf 34}, 470 (1986);  P. Pearle, Phys. Rev. A {\bf 39}, 2277 (1989);  G. C. Ghirardi, P. Pearle, and A. Rimini, Phys. Rev. A {\bf 42}, 78 (1990); P. Pearle, in {\em Quantum Theory: A Two-Time Success Story} (Yakir Aharonov Festschrift), eds. D. C. Struppa \& J. M. Tollakson (Springer, 2013), Chapter 9. [arXiv:1209.5082].  For a review, see A. Bassi and G. C. Ghirardi, Physics Reports {\bf 379}, 257 (2003).
\item S. Weinberg, paper in preparation.
\item This theorem was suggested to me by G. Moore, private communication.
\item E. P. Wigner, Ann. Math {\bf 40}, 149 (1939).
\item On this, see Banks, Susskind, and Peskin, ref. 8; G. C. Ghirardi, R. Grassi, and P. Pearle, Found. Phys. {\bf 20}, 1271 (1980).
\item P. Pearle, Eur. J. Phys. {\bf 33}, 805 (2012) [arXiv: 1204.2016].
\item K. Kraus, {\em States, Effects, and Operations -- Fundamental Notions of Quantum Mechanics}, Lecture Notes in Physics 190 (Springer-Verlag, Berlin, 1983): Chapter 3. 
\item W. F. Stinespring, Proc. Am. Math. Soc. {\bf 6}, 211 (1955); M. D. Choi, J. Math. {\bf 24}, 520 (1972).  For a review, see F. Benatti and R. Florentini,  Int.J.Mod.Phys. {\bf B19}, 3063 (2005) [arXiv:quant-ph/0507271].
\item M. D. Choi, Linear Algebra and its Applications {\bf 10}, 285 (1975).
 
\end{enumerate}

\end{document}